\documentclass{article}
\usepackage{amsfonts}  
\usepackage{amsthm}
\usepackage{amsmath}
\evensidemargin=\oddsidemargin
\begin{document}
\title{The Hamiltonians of Linear Quantum Fields:\\
I.  Existence Theory for Scalar Fields}
\author{Adam D. Helfer\\
Department of Mathematics\\ University of Missouri\\
Columbia, MO 65211, U.S.A.}
\maketitle
\begin{abstract}
For linear scalar field theories, I characterize those classical
Ham\-il\-tonian vector fields which have self-adjoint  operators as their
quantum counterparts. As an application, it is shown that for a scalar field in
curved space--time (in a Hadamard representation), a self-adjoint Hamiltonian
for evolution along the unit timelike normal to a Cauchy surface exists only if
the second fundamental form of the surface vanishes identically.
\end{abstract}

\newtheorem*{thmm}{Theorem}
\newtheorem*{corr}{Corollary}
\newtheorem{proposition}{Proposition}
\newtheorem{definition}{Definition}
\newtheorem{corollary}{Corollary}
\newtheorem{theorem}{Theorem}
\newtheorem{lemma}{Lemma}

\def\tr{{\rm tr}}
\def\rSp{{\rm Sp}_{\rm rest}{}}
\def\Sp{{\rm Sp}}
\def\i{{\rm i}}
\def\d{d}        
\def\e{e}        
\def\cA{{\mathcal A}}
\def\C{{\mathbb C}}
\def\R{{\mathbb R}}
\def\A{{\mathbb A}}
\def\Z{{\widehat Z}}
\def\wp{{\widehat\phi}}
\def\wH{{\widehat H}}
\def\a{{\widehat a}}
\def\H{{\wH}}
\def\X{{\hat X}}
\def\wpsi{{\hat\psi}}
\def\DA{{{\cal D}(A)}}

\section{Introduction\label{Introduction}}

When does a symmetry of a classical field theory pass over to the
corresponding quantum theory?  

Even for linear theories, the answer to this question is not known in
the generality one would like.  For {\em finite} motions, indeed, the
answer is well-known:  a canonical transformation of the classical
phase space induces a Bogoliubov transformation on the field operators,
and this transformation is unitarily implementable if and only if the
Bogoliubov coefficients satisfy a certain square-summability condition
(Shale 1962).  However, the most important canonical transformations
are those corresponding to evolution in time.  For these, one almost
never has an explicit knowledge of the finite transformations ---
getting these would involve solving the equations of motion.  What one
has is knowledge of the {\em generator} of the transformations --- the
equation of motion, or equivalently the classical Hamiltonian function
or vector field.  One would like to be able to read off from this
whether or not the quantum evolution will be unitarily implementable.
It is this problem which is solved, for scalar fields, in this paper.
The analysis of other bose fields is parallel.  Subsequent papers in
this series will explore the structure of the Hamiltonians more fully,
and treat fermions.

In the past few years, it has become apparent that this issue is
important, because evidence has accumulated which strongly suggests
that in generic circumstances the Hamiltonians are {\em not }
self-adjointly implementable.  Indeed, this appears to be one of a
family of related phenomena, which at least superficially are severely
pathological.  They are all local, and can be expressed as
certain ultraviolet divergences.

The most extreme of these phenomena is that, generically, the
expectations of the energy and of the energy density
are unbounded below.  
Schematically,
\begin{equation}
 \mbox{lower bound of }\H =(\mbox{finite term})+(\mbox{geometric term})
(-\infty )\, ,
\end{equation}
where the ``geometric term'' vanishes in Minkowski space but is
generically non-zero.  What this means is that, at least as
far as the mathematical structure of the theory is concerned, the case
of evolution along a covariantly constant timelike vector field in
Minkowski space is a highly unstable point.

It would be hard to overstate the potential significance of this issue.  It
raises fundamental questions of stability and interpretation.  For example, why
do not perturbations send the quantum field cascading through more and more
negatively energetic states, with a corresponding release of positive-energy
radiation?  The expectation of the energy density is used as a source term in
``semiclassical gravity,'' which is perhaps the most important application of
quantum field theory in curved space--time.  (It is this theory, for example,
which predicts the loss of energy from black holes via the Hawking mechanism.) 
How credible is this semiclassical approximation, in view of the
unboundedness-below?

These matters are at present imperfectly understood.  There is a plausible
resolution for them in the case of special-relativistic quantum fields, but in
the case of quantum fields in curved space--time, if a similar picture is to
hold, we seem to have to confront quantum fluctuations in the geometry in an
essential way.  I shall outline this picture below.  It must be emphasized,
though, that at present, even in the special-relativistic case, the picture is
one of physical plausibility.  To justify (or negate) it, one needs a firm
understanding of what the mathematical structure is.  The main aim of the
present series of papers is to lay the foundations of the general theory of the
Hamiltonians involved.

%
%

\paragraph{The Emerging Picture}

Workers in the theory of quantum fields in time-dependent external potentials
have been aware of some of the issues raised here for some time, but the ideas
have been surprising to workers in quantum field theory in curved space--time. 
I should like to explain here what the significance of these issues is, and in
particular why, for quantum fields in curved space--time, they may be of the
deepest significance.  I will also try to give a modern point of view of the
state of affairs for  special-relativistic quantum fields in time-dependent
external potentials. 

Of course, any attempt to say what picture is ``emerging'' before it
has actually emerged involves some judgments, which may not be shared
with all workers.  I shall comment in particular on the relation
between the view here the ``algebraic
approach,'' below.

\paragraph{Failure of Unitary Implementability}

I shall first discuss the non-unitari\-ties, that is, the failure of evolution
of the field for finite times to be unitarily
implementable.\footnote{Throughout these papers, we use the conventional
``relativistic Heisenberg'' picture, in which the state vector is unchanged
(except when reduction occurs) and the field operators evolve with time.  Thus
probability is automatically conserved.  The problematic unitarity concept is
unitary implementability of evolution.  This is explained in a little more
detail in section~\ref{bolivar}}  While the present papers are really
concerned with the Hamiltonians (that is, evolution for infinitesimal time),
one should understand the significance of the non-unitarities first.  Such an
understanding has only gradually developed in the case of special--relativistic
field theories, and to my knowledge it has not been discussed in the case of
quantum fields in curved space--time.

Consider for definiteness the case of quantum charged particles responding to
an external, classical, electromagnetic field.  It is often asserted that,
while there may be ``intermediate'' non-unitarities in this case (that is, if
one only considers evolution to some intermediate time), they are not
significant, because the S--matrix  is unitary.

This position is indeed tenable if all one is interested in is the S--matrix,
and only in those cases when it does turn out to be unitary --- typically
scattering through a potential which is turned on and off in finite time.  But
if one is interested in the intermediate regime, this is not a satisfactory
position. When we come to quantum fields in curved space--time, we almost never
``switch on and off'' the gravitational field in finite time, so we must face
the intermediate regime.\footnote{Also the position is not satisfactory for
investigations of quantum measurement issues, where one wishes to take into
account the facts that observers investigate only finite volumes of
space--time, and that different observers can never synchronize their frames
perfectly.  The unitarity of the S--matrix only applies to evolution from one
complete $t=$ constant surface in Minkowski space to another $t'=$ constant one,
where both $t$ and $t'$ are inertial coordinates.  Thus if data are taken over
a surface by several observers who have uncertainties in synchronizing their
clocks, the unitarity fails (Helfer 1996).}

What happens in this intermediate regime?  In special-relativistic quantum
field theories, presumably the non--unitarity results from neglecting
fluctuations in  the external potential; when these fluctuations are included,
the evolution is  supposed to be unitary.\footnote{Of course, we do not know
that the evolution really is unitary, because we can only treat this theory ---
full, nonlinear, quantum electrodynamics in the case of charged particles and
electromagnetic fields --- perturbatively. Serious workers have occasionally
suggested dropping the unitarity requirement as a way of evading the
conclusions of Haag's theorem.} This is indeed a plausible picture, and if
correct explains away the non-unitarities in special-relativistic problems. 
Since as we shall see in these papers, the non-unitarities are closely linked
to the unboundedness--below, it  is also plausible (but not a foregone
conclusion) that accounting for the fluctuations will semi-bound the
Hamiltonians.

However, there is more to the story.  In order to understand this,
recall that for linear field theories the non-unitarities are determined by
when the Bogoliubov coefficients fail to satisfy a square-summability
condition.  In the most general situations, the sum involved may diverge because
of:  (a) infrared problems; (b) ultraviolet problems; (c) resonance phenomena
assigning divergent weights to moderate modes.  In what follows, I shall
only consider ultraviolet divergences, but similar comments could be made for
the other cases.

If the evolution fails to be unitarily implementable on account of ultraviolet
divergences, then, with any finite ultraviolet cutoff $\Lambda$, the evolution
is in fact unitarily implementable.  What does this mean physically?  

The answer to this can be found by examining the formulae for the quantum
evolution operators in terms of the Bogoliubov transformations (e.g., in
section~\ref{Existence}, below).  Details of this will appear elsewhere; here I
shall only indicate the results. One examines how the evolution operator
$U_\Lambda$ varies with the cutoff $\Lambda$, in particular, how $U_\Lambda$
affects modes of different frequencies.  One finds that there is a critical
scale $\Lambda _0$ which can be estimated in terms correlations of the external
potential.   For $\Lambda\ll\Lambda _0$, the operator $U_\Lambda$ varies
smoothly with $\Lambda$. For $\Lambda \gg\Lambda _0$, the restriction of
$U_\Lambda$ to the low-frequency sector has a reasonably well-defined limit;
however, the action on high-frequency modes varies essentially chaotically (in
a loose sense) as $\Lambda$ is increased.

What this means is that, in the cut-off theory but for $\Lambda\gg\Lambda _0$,
qualitatively new features appear in the evolution.  While these are not,
strictly speaking, non-unitarities, they are certainly interesting.

At the moment, it is not known whether this ``chaotic'' behavior actually
occurs in any physically realizable situations.   If there is a scale $\Lambda
_{\rm fluct}$ at which fluctuations in the external field must be accounted
for, then the question is whether $\Lambda _0$ can exceed $\Lambda _{\rm
fluct}$.  If not, then the fluctuations in the field become the dominant
feature before the chaotic regime can be reached; on the other hand, if it is
possible to have $\Lambda _0<\Lambda _{\rm fluct}$, then there can be a regime 
$\Lambda _0<\Lambda <\Lambda _{\rm fluct}$ in which effectively chaotic
behavior occurs before the external-field approximation must be superseded. It
may be that as a matter of principle this last possibility is excluded, but at
the moment this remains an open question. It seems most likely that insight
will be gained by investigating specific physical models.

\paragraph{The Unitarity Problem for Quantum Fields in Curved Space--Time}

But the most important application of the present analysis of intermediate
non-unitarity is to quantum fields in curved space--time.  In this case, it is
the \it gravitational \rm field which is the external potential.  If a picture
like that expected for special-relativistic theories applies, then \it
unitarity is restored by fluctuations in the gravitational field.  \rm  It
should be possible to build on the present analysis to make precise statements
about what the characters of these fluctuations must be.  In other words, we
may be able to use the present work to \it deduce \rm  quantum characteristics
of the gravitational field, assuming unitarity still holds in that context. It
should be emphasized that these issues arise at moderate scales, far below the
Planck threshold.  Thus we have a new line of attack on quantum gravity, one
which does not require hypotheses about the strong-gravity, Planck-scale,
regime.

\paragraph{Relation to Algebraic Approaches}

It has been recognized for a long time that it is useful to regard
quantum field theory as constructed in two stages.  First, one
describes precisely an algebra $\A$ of observables; then, one seeks a
representation of this algebra on a physically acceptable Hilbert
space.  Specifying the algebra $\A$ is rather like specifying a group by giving
its multiplication table;
specifying the representation is realizing the
observables as operators on a Hilbert space.

Generally, by an {\em algebraic approach} to quantum field theory, one
means an approach which starts from and emphasizes the algebra $\A$.
The most extreme form of this approach would seek to describe physics
entirely in terms of this algebra without the construction of a
representation.  More moderate approaches (which are common) do make
use of representations, although these representations are analogs of
density matrices rather than state vectors.  (The present paper, which
emphasizes the construction of satisfactory Hamiltonians acting on
Hilbert spaces of state vectors, would not be considered an algebraic
approach.)  One can recast quantum field theory in such algebraic
terms, and then one has a formalism equivalent to the conventional
one, and so it becomes a matter of taste or convenience which is to be
preferred.  

Algebraic approaches are attractive for some purposes, because they
treat different representations with equal facility.  In particular,
when quantities which are singular in one representation may have 
sensible existences in other representations (such as the candidates 
for Hamiltonians in these papers), the algebraic approach can be a 
natural way of accommodating them.  Indeed, the Hamiltonians have 
natural algebraic existences, because they are generators of symmetries
of the field algebra $\A$.

However, precisely this feature --- the algebraic approach's
egalitarian treatment of different representations --- can be a
drawback.  Precisely because of this, it is difficult or impossible to
see, from the algebraic structure alone, how physically preferred
representations (such as the ones of interest here) are
distinguished.  

While the results of these papers could be cast in terms of the
algebraic approach, the papers' goals are more naturally met in a
conventional, representation-based, approach.  While the present work
is very mathematical, its goals are to provide enough structural
background that it will be possible for subsequent work to develop a
physical understanding of the significances of non-unitarities and
ill-defined Hamiltonians.  These are very much representation-based
concerns, with the physically-dictated choices of representations
playing a central role.  Thus the representational approach seems to
focus most directly on the physical issues.

I have emphasized that the point of the present papers is to lay a
framework within which the significance of some potentially serious
pathologies (non-unitarity, unbound\-edness-below of energy) can be
analyzed, and I have argued that the representation-based approach is
probably more suited to this than an algebraic one.  On the other
hand, once the physical significance of the pathologies is understood,
one may have to reassess what the most appropriate formalism for
the theory is.  If, for example, it were to turn out that the
pathologies were not of much physical significance, but only somehow
mathematical niceties, then it could well be that the algebraic
approach would be most useful, as a framework which can accommodate
such behaviors without giving them undue weight.

\paragraph{The Hamiltonian}

I shall now discuss the problematic nature of the Hamiltonian.

Of course, the Hamiltonian is simply the derivative of the evolution operator,
so all of the comments above should apply, interpreted ``infinitesimally,'' to
the Hamiltonian.  However, to have such an interpretation, one needs an
infinitesimal counterpart of the Bogoliubov--Shale criterion for unitary
implementability,  that is, a criterion for the existence of the Hamiltonian as
a self-adjoint operator.  Such criteria are provided in these papers. We may
expect these criteria to be more practically useful, for many purposes, than
that of Bogoliubov and Shale, since criteria in terms of the Hamiltonian do not
require integration of the equations of motion.  

Assuming the picture suggested above  (of unitarity being saved by the
breakdown of the validity of the external-field picture) for
special-relativistic theories is correct, one would expect that the problems
of ill-definition (as a self--adjoint operator) of the Hamiltonian are
resolved by the same breakdown. Parallel to the question of the existence of an
``effectively non-unitary regime,'' one  has the question of whether there is
a sort of ``effectively non-self-adjoint regime.''  In other words, in a
theory with cutoff $\Lambda$, the Hamiltonian will depend sensitively on
$\Lambda$ for $\Lambda\gg\Lambda _0$.  If $\Lambda _0$ is less than the scale
$\Lambda _{\rm fluct}$ at which the external-field approximation breaks down,
then one does have such a regime.  Developments of the techniques of this paper
should enable one to estimate $\Lambda _0$ in terms of the external field.

The foregoing applies to the Hamiltonian as the generator of evolution. 
However, there are important points beyond this.  These have to do with the
interpretation of the Hamiltonian as an energy operator, and with negative
energies.  

We wish to consider not just the total Hamiltonian, but the various energy
operators that can be formed from the stress--energy, by integrating it against
a smooth test field on a four-volume.  Physically, these correspond to
measurements of energy (or momentum, angular momentum, etc.) in finite spatial
volumes with temporal averagings as well.  These operators presumably are what
figure in real experiments, since one can never measure a total (over all
space) energy.  The picture that emerges from these papers, as well as from
previous work, is that (in the case of averaged energy operators), is that such
averaged Hamiltonians exist as semi-bounded self-adjoint operators.  But as
the temporal extent of the averaging is decreased:  (a) the lower bounds
diverge to $-\infty$; and (b) the Hilbert-space domains of the operators become
more sparse, having only $\{ 0\}$ as a limit.

It should be emphasized that this behavior occurs even in the simplest cases,
for instance, for a Klein--Gordon field in Minkowski space (with no external
potential).  Consider for definiteness such a field, and an averaged energy
operator
\begin{equation}
{\widehat H} (b,B)=\int {\widehat T}^{\rm ren}_{tt} (t,x,y,z)\, b(t)
  B(x,y,z) \, \d t\d x\d y\d z\, ,
\end{equation}
where $b(t)\geq 0$, $B(x,y,z)\geq 0$ are smooth 
compactly-supported test functions effecting the 
temporal and
spatial smearing of the energy density operator; we require
\begin{equation}
 \int b(t)\, \d t=1
\end{equation}
to treat $b$ as an averaging.
For any fixed $b$, $B$, the
quantity ${\widehat H}(b,B)$ is a self-adjoint operator, bounded below.  But as
the function $b$ becomes more sharply peaked at a given time, we find the
conditions (a) and (b) of the previous paragraph.  This means that there is no
idealized operator
\begin{equation}
  \lim _{b(t)\to\delta (t)} {\widehat H}(b,B)
\end{equation}
representing the measure of the energy in a finite three-volume.  In other
words, ``the energy in a spatial volume'' is not a meaningful quantum
observable.  It is necessary to specify the scale of the temporal averaging in
order to have a well-defined self-adjoint operator.

\paragraph{Semiclassical Gravity}

While many workers have recognized this point, it has not been taken  fully to
heart, I think, in an important area:  attempts to understand the
back-reaction of the quantum field on space--time.  One would expect that this
is determined by the energy density operator (and more generally by the
components of the stress--energy), but we have just seen that this operator
does not really exist; only averaged versions exist.  This means that one must
specify the scale of the averaging before one has a well-defined theory. In
other words, the specification of the averaging scale is not just a
technicality which will sort itself out; different scales give different
theories, and one needs to know if it is possible to identify a physically
plausible one for a given problem.

To be explicit, let us consider the most common starting-point for treatments
of the back-reaction of the quantum field on space--time, the
``semiclassical approximation:''
\begin{equation}
  R_{ab}-(1/2)Rg_{ab}=-8\pi G\left( T^{\rm classical}_{ab} +
    \langle {\widehat T}^{\rm ren}_{ab}\rangle\right)\, .
\label{tuckwell}
\end{equation}
This equation is a natural one to write down, but is it justified?  

In general, one would expect such an approximation, where the quantum fields
only couple through expectation values of an operator $\widehat O$, to be valid
when fluctuations in $\widehat O$ are negligible.   It is easy to construct
examples, even in Newtonian gravity and non-relativistic quantum mechanics, in
which this fails. Consider for instance two separated boxes, one at ${\bf
r}={\bf a}$ and the other at ${\bf r}=-{\bf a}$, and a particle of mass $m$
with probability $1/2$ of being in either box.  Assuming the sizes of the boxes
are negligible (compared to $\|{\bf a}\|$), then the gravitational potential
constructed from the expectation of the energy density is
\begin{equation}
  -{\frac{Gm/2}{\| {\bf r}-{\bf a}\|}} -{\frac{Gm/2}{\| {\bf r}+{\bf a}\|}}\, ,
  \label{ulrich}
\end{equation}
whereas the correct potential is presumably a quantum superposition
\begin{equation}
{\frac{1}{\sqrt{2}}}
\left| -{{Gm}\over{\| {\bf r}-{\bf a}\|}}\right\rangle\ +{\frac{1}{\sqrt{2}}}
  \left| -{{Gm}\over{\| {\bf r}+{\bf a}\|}}\right\rangle\, .
  \label{veronica}
\end{equation}
These two states of the gravitational field --- (\ref{ulrich}) and
(\ref{veronica}) --- are different, and that difference is in principle
detectable in many ways.  For example, an observer near one box experiences a
certain gravitational red-shift relative to infinity
according to (\ref{ulrich}); according to
(\ref{veronica}) she experiences, with probability $1/2$, either a much smaller
or an almost doubled red-shift.\footnote{While the argument in this
paragraph has been deliberately phrased to make it as homely and
unprovocative as possible, it implicitly involves
quantum--gravitational assumptions.  For example, the notion that the
potential can be in a superposition of states implies that the
potential must really be a quantum observable.}

What this example shows is that in principle it is quite possible for the
semiclassical approximation to fail, even in mild circumstances.  Whether this
failure is significant or not depends on exactly what regime one hopes to apply
the approximation to.  It is presumably far beyond the state of present
technology to detect the difference between (\ref{ulrich}) and (\ref{veronica})
with a laboratory experiment --- one would need to be able to measure very fine
differences in red-shifts with apparatuses much smaller than the scale
$2\|{\bf a}\|$ of the separation of the position eigenstates.  

In general, for any proposed application of the semiclassical approximation, in
order to check its credibility one needs to:  (a) estimate the fluctuations;
(b) specify the accuracy to which space--time is supposed to be modeled; (c)
specify the time for which that model is supposed to be valid, before the
errors made in the approximation accumulate and become significant.  In some
circumstances, such as the non-relativistic example above, elementary
estimates can quickly convince one that the approximation is valid for
practical purposes.  However in other circumstances this is may not be so
obvious.

Let us return now to the case of quantum fields in curved space--time.  It is
not hard to show that as the temporal averaging tends to zero, the fluctuations
in $\widehat O$ (here $\widehat O$ is a temporally-averaged ${\widehat T}^{\rm
ren}_{ab}$)  not only dominate the expectation value but actually diverge for
{\em all } Hadamard states.\footnote{This follows from results in Helfer
(1996).  In terms of the notation of equation (\ref{aloysius}), below, the
fluctuations contain a term $C_{\alpha\beta}{\overline C}^{\alpha\beta}$, which
diverges.} This clearly raises questions about the validity of the
semiclassical approximation.  It is presumably possible to answer these in many
cases; it is not possible to ignore the questions.

The divergent fluctuations of ${\widehat T}^{\rm ren}_{ab}$ mean that equation
(\ref{tuckwell}) is not credible if interpreted literally. The equation might
be valid if the right-hand side were replaced by some sort of temporal (or
space--time) average of the expectation of the stress--energy.  This indeed is
presumably what happens in most everyday circumstances, where the quantum
character of the matter is negligible for the purposes of understanding its
effect on space--time.  However, the main interest of quantum field theory in
curved space--time attaches to those situations in which the quantum character
of the fields is important.  In those circumstances it becomes problematic to
justify an equation like (\ref{tuckwell}), even with a re-interpreted
right-hand side.  To do this, one would have to spell out the accuracy to
which one wanted to model space--time classically, the time for which that
model should hold, and the scale of the averaging.  One would have to justify
physically the averaging procedure.  These sorts of questions have been very
little investigated (see however Flanagan and Wald 1996 and Ford and Wu
1999).  

Whether the semiclassical approximation turns out to be valid for a given
purpose or not, careful attempts at its justification will deepen our
understanding of the physics of the situation, as we ask exactly what scales
are important, and how do we expect the behavior of the field to affect the
space--time.   And certainly the most interesting cases would be those for
which (\ref{tuckwell}) did not hold, even with a re-interpreted right-hand
side.  In such cases one would have to take into account the quantum character
of the gravitational field. It is quite possible that such situations do
exist:  above, we saw that it seems necessary to include quantum fluctuations
in the gravitational field in order to restore the unitarity of the field
theory.  

\paragraph{Negative Energies}

Much of the interest in these quantum stress-energies is on account of the
prediction of relativistic quantum field theories that the renormalized,
temporally averaged, energy densities can be negative.  Thus if these
stress-energies can be used as sources for Einstein's equation, one has a
violation of the Weak Energy Hypothesis, a key assumption underlying some of the
most important results in classical relativity (the singularity theorems and the
Area Theorem).  Thus the understanding of in precisely what sense the energy
densities can be negative is a key problem, on whose resolution depend basic
qualitative aspects of the behavior of space--time.

We must bear in mind that at present there is
no direct experimental evidence of negative energy
densities.\footnote{The measurement of the Casimir force is sometimes cited as
such evidence, but that is a measurement of one of the ``pressure'' components
of the stress--energy, not the energy density.  There is only an indirect link
between the two operators. Moreover, there are two sorts of concerns about the
usual prediction of negative energy density between the plates of the Casimir
apparatus.  The first is that the plates have been unphysically idealized as
perfect conductors; rough estimates show that for real plates there could be a
separation-independent positive contribution to the energy density
overwhelming the usual negative energy density (Helfer and Lang 1999);
this would not alter the predicted Casimir force.  Second,
there may be difficulties of quantum measurement theory in meaningfully
ascribing a negative energy density to the region between the plates (Helfer
1998).} Thus the importance of having as complete a theoretical understanding
of these as possible.

\paragraph{Formal Structures}

To explain the foregoing more quantitatively,
let us begin by writing down the formal expression for a
Hamiltonian operator for a linear theory:
\begin{equation}
 \H =C^{\alpha\beta} {\a}_\alpha {\a}_\beta
  +B^\alpha{}_\beta {\a}^{*\beta} {\a}_\alpha +{\overline
C}_{\alpha\beta} {\a}^{*\alpha} {\a}^{*\beta} +\mbox{c-number term}\, .
  \label{aloysius}
\end{equation}
Here $\a$, $\a ^*$ are annihilation and creation operators, and
$B^\alpha{}_\beta$, $C^{\alpha\beta}$ are coefficients
(self-adjoint and symmetric, respectively).\footnote{The
modes annihilated and created by $\a$, ${\a}^*$ need not be particle
modes; this will be discussed further, below.
Also, the
formal Hamiltonian~(\ref{aloysius}) has been written 
in normal-ordered form.  This is
for purposes of orientation only.  In the analysis that follows,
criteria are developed for determining whether the quantum Hamiltonian
exists without prior assumptions about what renormalization scheme is
to be used.  The second paper in this series will contain some further
results, about when normal-ordering is adequate to define the theory.}
If $C^{\alpha\beta}=0$
and $B^\alpha{}_\beta$ is self-adjoint, then the definition of the
Hamiltonian is unproblematic; one has a structure much like that of the
Hamiltonian for a free Klein--Gordon field in Minkowski space.
However, if this situation is perturbed even slightly, difficulties may
appear.  
For example, it is easy to see that the vacuum $|0\rangle$
cannot be in the domain of $\H$ unless $C^{\alpha\beta}{\overline
C}_{\alpha\beta}<\infty$.  This condition can very well be violated,
even if the $C^{\alpha\beta}$ are uniformly small.  

More severe problems may occur.  It may be impossible to find {\em any
} normalizable states on which $\H$ is well-defined.  This means that
$\H$ can have only a very limited existence, and certainly cannot be
the self-adjoint operator that quantum theory requires an observable
to be.  These sorts of difficulties are potentially very serious, as
they raise the question of what the operator character of the
Hamiltonian, and hence of the dynamics, is.  This should be contrasted
with somewhat finer issues of renormalization which go to the kinematic
question of what the c-number contribution to $\H$ is, for example,
the computation of Casimir or ground-state energies.  Those finer
issues presumably cannot even be addressed until the operator character
is properly understood.

We have seen that difficulties are potentially present when
$C^{\alpha\beta}\not= 0$.  Physically, this indicates that evolution by
the Hamiltonian does not preserve the decomposition of the field into
creation and annihilation parts.  This situation occurs naturally in
many settings.  Most obviously, it is the situation in time-dependent
external potential problems.  In particular, it is the {\em generic}
case for quantum fields in curved space--time.  However, it can also
occur when there is no explicit time dependence in the theory.  If one
has several linearly coupled fields, for example, in general one has
$C^{\alpha\beta}\not=0$; cases of of current interest are the models
for the quantum electromagnetic field in dispersive dielectric media
(see, e.g., Barnett 1997).  And the theory of squeezing revolves
precisely around Hamiltonians with $C^{\alpha\beta}\not=0$ (Loudon and
Knight 1987).

\paragraph{Nature of the Present Paper}

The main theorems to be given here have characters similar to some
basic results in quantum theory, in that their general
physical content can be appreciated without pursuing the analytic
technicalities of their proofs.  (Thus for example, physicists use daily
spectral resolutions of self-adjoint operators without worrying about how
the existence of such resolutions is proved; and one can appreciate the
sense of the Stone--von~Neumann theorem, that canonical quantizations of
mechanical systems with finitely many degrees of freedom are unitarily
equivalent, without examining its proof.)  I have written these papers so
as to confine the analytic technicalities to the proofs.  I hope that
the statements of the theorems will be accessible to general workers in
quantum field theory.

There has been some previous work in this area.  As mentioned earlier,
Shale (1962) found the condition for a {\em finite } evolution to be
unitarily implementable.  Analyses parallel to Shale's are central to
the theory of loop groups, and it is possible that the present results
may have analogs of interest there.  Klein (1973) investigated the case
of Hamiltonians, and provided a host of counterexamples to 
natural conjectures.  More recently, Honegger \& Rieckers (1996)
established some results under fairly strong hypotheses on
$C^{\alpha\beta}$.

The plan of the paper is this.  The next section contains a brief
discussion of the concept of unitary implementability.  This can be
skipped by those who understand the distinction between this sort of
unitarity and that governing the state vector.
Section~\ref{Preliminaries} contains preliminaries, mainly the basic
definitions needed to state the problem mathematically.
Section~\ref{Characterization} proves one of the two main theorems
(Theorem~\ref{hermione}), characterizing which classical Hamiltonian
vector fields give rise to one-parameter unitary groups on the
physical Hilbert space.  In Section~\ref{Existence}
(Theorem~\ref{guillermo}), it is shown that each group that arises in
this way is automatically strongly continuous, and so possesses a
self-adjoint generator, that is, a quantum Hamiltonian.
Section~\ref{Examples} gives as an example the case of quantum fields
in curved space--time; it is shown (Theorem~\ref{Schubert}) that
evolution along the timelike unit normal to a Cauchy surface is not
self-adjointly implementable unless the second fundamental form of the
surface vanishes identically.  Section~\ref{Summary} contains
some comments.

The assumptions of the present paper are very general.  In the next
paper, I specialize to the case where the classical Hamiltonian
functions are positive.  Then one can say much more about the structure
of the theory, and take up the question of whether the quantum
Hamiltonians are bounded below.  The third paper in the series will
treat fermions.

{\em Background.}
A good summary of the necessary quantum field theory, from the point of
view of this paper, will be found in Wald's (1994) book.  The functional
analysis can be found in Dunford and Schwartz (1958).  In Section 5, I
have made use of the theories of pseudodifferential and Fourier integral
operators, for which see Treves (1980).

{\em Summary of Notation.}  Here is a summary of the notation used. 
Unfortunately, there are quite a few things denoted conventionally by
similar symbols.

$H$ is the space of solutions of the classical field equations, a real
Hil\-bert\-able space equipped with a symplectic form $\omega$.

$H_\C$ is the space $H$ equipped with the complex structure defined by
$J$, and so made into a complex Hilbert space.

${\cal H}$ is the physical Hilbert space 
of the quantum field theory, that is, the space on which the representation 
of the field algebra acts.

$\|\cdot\| _{\rm op}$ is the operator norm.

$\|\cdot\| _{\rm HS}$ is the Hilbert--Schmidt norm.

$\A$ is the field algebra.

$A$ is the Hamiltonian vector field on the space of classical solutions.

$\cA$ is the Lie adjoint of $A$, that is, the derivative of conjugation 
by $g(t)=e^{tA}$.

{\em Notes.}  Since $H$ has no preferred inner product, I have usually
been careful to emphasize the dependence of properties on $J$.  Thus one
has $J$-symmetric transformations, etc.
The Hilbertable and Hilbert spaces used here are always
assumed to be separable, that is, to have countable bases.

\section{Unitary Implementability\label{bolivar}}

The central question in this paper is, When does a group of motions on
classical phase space have a unitary counterpart on Hilbert space? 
Since most of the quantum field-theoretic literature does not
distinguish explicitly between this sort of unitarity and that
governing the evolution of the state vector, it seems worthwhile to
spell this out.

Let us consider a linear quantum field theory in
the presence of a perhaps time-dependent external potential.  This is
constructed
in two steps.  First, one defines an {\em algebra of fields}
$\A$.  These are not yet field {\em operators}, as they do not as yet
operate on anything.  Rather, the algebra $\A$ is a mathematically
precise way of expressing the canonical commutation relations which
{\em any} such operators will be required to have.  The second stage of
the construction is the identification of the fields with specific
operators on a Hilbert space, that is, the specification of a {\em
representation} of the algebra $\A$.  (One can think of the steps as
analogous to first defining a group by a multiplication table, and then
giving a realization of it as a set of matrices acting on
column-vectors.) In a linear field theory, the algebra $\A$ is
essentially determined by the classical phase space of the theory,
since the canonical quantization specifies the commutation relations in
terms of the Poisson brackets.  There are in general many inequivalent
representations one might choose for this algebra, and the question of
which one is physically correct may be subtle. 

For quantum fields in curved space--time, we accept the standard point of view,
that the correct representation is a ``Hadamard'' one (cf. Wald 1994).  This
can be specified in various equivalent ways, for example, by demanding that on
a dense set of states the leading short-distance asymptotics of the two-point
functions agree with those in Minkowski space.  However, it is only in
section~\ref{Examples} that the details of the Hadamard form are used.  Because
the rest of this paper has a very general mathematical perspective, the
conclusions depend only on the fact that the physics {\em does } determine a
representation.\footnote{More precisely, the Hadamard condition
determines a certain class, or ``folium,'' of representations.}
Thus most of the results are phrased in terms of compatibility
issues between a generator of symmetries of $\A$ (which we wish to implement as
a quantum operator) and a representation; the precise way the representation is
determined figures only in section~\ref{Examples}.   (The general class of
representations we are concerned with are those given by symplectic
quantization; explicit formulae are given in
section~\ref{Existence}.\ref{Representation}.  For  particulars of the Hadamard
representations, see Wald 1994; we also use formulae from Helfer 1996.)

In brief, then, besides the canonical commutation relations and the
field equations, one needs an extra input to construct a quantum field
theory:  the choice of representation.  In Minkowski space, in the
absence of fixed external fields or boundary conditions which might
break the relativistic invariance, one can find an essentially unique
Poincar\'e{}-invariant representation, the Fock representation.
However, in more general circumstances it can be a subtle issue to
determine the physically correct representation.  While we shall not
need this here, it may be remarked that the choice of representation is
encoded in the (infrared and ultraviolet) asymptotics of the two-point
functions.\footnote{And in any additional singularities the two-point
functions may have, for example on account of boundary conditions.}
Thus different representations may lead to different local quantum
fluctuations, different vacuum polarizabilities, etc.

All representations considered here will have the same abstract
mathematical form as the Fock representation in that they will be
determined by a decomposition of the field operator into ``creation''
and ``annihilation'' parts, with a corresponding ``vacuum'' state. 
However, the modes created and annihilated may not correspond to
particles, and may have no simple physical interpretations.  Likewise,
the ``vacuum'' state need not be interpretable as a physical vacuum. 
Such representations are adequate for almost all purposes, and more
general ones can be constructed as direct sums or integrals of these.

By the evolution of the fields, we mean their change when the classical
phase space is evolved along some Hamiltonian vector field, which in
our case shall always respect the linear structure of the phase space.
This evidently will determine an automorphism of the algebra $\A$, and
one would like to identify the generator of that automorphism with the
quantum Hamiltonian.  However, it may happen that the automorphism is
{\em not} induced by any unitary motions of the physical Hilbert
space.  For a one-parameter group of motions, this means that {\em the
Hamiltonian cannot be realized as a self-adjoint operator}.

Two points should be emphasized about this sort of non--unitarity:

\begin{itemize}

\item
The evolution in question is that of the
algebra of fields, and not that of the state vectors.  The state vectors
do evolve unitarily --- in fact, are unchanged in our
Heisenberg picture (except when reduction occurs).  

\item
The possibility of non-unitarily
implementable evolution occurs only in quantum {\em field} theory.  In
quantum {\em mechanics}, when there are only finitely many degrees of
freedom, the Stone--von~Neumann Theorem guarantees that any two
representations of the canonical commutation relations are unitarily
equivalent.  A corollary of this is that the Hamiltonians in the case
of fields must always be {\em formally} self-adjoint, in a suitable
sense.  For any ``coarse graining'' of a quantum field theory to
finitely many degrees of freedom will result in evolutions which are
unitarily implementable.  This means that the failure of unitary
implementability, or of self-adjointness, must occur in the passage
to the limit of infinitely many degrees of freedom.

\end{itemize}

\section{Preliminaries\label{Preliminaries}}

I shall begin by indicating, for the non-experts, the meanings of the
objects in the symplectic treatment of the quantization of linear bose
fields.  Those familiar with this can skip to Section 2.2 to identify
the terminology and symbols used in this paper.

\subsection{Orientation}

The usual Fock construction of free field theories in Minkowski space
can be generalized to apply to linear fields responding to external
potentials, in particular to fields in curved space--time.  I shall
summarize here how this is done.

Let $H$ be a real Hilbertable\footnote{A {\em Hilbertable} space is a
topological vector space whose topological structure can be determined
by an inner product, but without a preferred choice of inner product.
The restriction to Hilbertable spaces is made to streamline some of the
analysis, and could be weakened.} space, which in applications is the
space of solutions (of a certain Sobolev regularity) to the classical
field equations for a Bose field.\footnote{In the case of a gauge
field, the following treatment applies to gauge equivalence classes of
solutions.}  This may be either in flat or curved space--time, and
external fields may be present.  We shall not explicitly discuss
charged fields, but these can be treated with straightforward
modifications of the present techniques.  We require that there be
given a symplectic form $\omega$ on $H$.  Then $\omega$ determines an
abstract algebra $\A$ of fields, obeying the canonical commutation
relations.  The term ``abstract'' is used here to emphasize that there
has been as yet no construction of the quantum Hilbert space and
representation of the algebra as operators on the space.

A particular representation 
(of the sort usually considered)
of the field algebra is determined by
choosing a {\em positive complex structure}, that is a map $J:H\to H$
which preserves $\omega$, satisfies $J^2=-1$, and such that $\omega
(v,Jv)$ is positive-definite.  For free fields in Minkowski space, one
chooses $J\phi =\pm \i (\phi _+-\phi _-)$, where $\phi _\pm$ are the
positive- and negative-frequency parts of $\phi$.  In Minkowski space,
then, the positive-frequency fields are the $+\i$ eigenspace of $J$,
and from these the Fock representation is constructed in the usual way.
The same mathematical prescription for constructing a representation of
$\A$ works, however, for {\em any} positive complex structure $J$ on any
Hilbertable symplectic space.

The choice of $J$ is physically important.  Different choices of $J$
will generally lead to inequivalent representations of $\A$:

\begin{thmm}{(Shale 1962)}
Two positive complex structures, $J_1$ and $J_2$, lead to
unitarily equivalent representations of $\A$ iff $J_1-J_2$ is
Hilbert--Schmidt.
\end{thmm}

\noindent (Recall that an operator $L$ is Hilbert--Schmidt if ${\rm
tr}\, L^*L<\infty$.  Note that in finite dimensions, all operators are
Hilbert--Schmidt.)  It turns out that, at least for linear scalar
fields in curved space--time, there is a natural choice of $J$, or more
properly, an equivalence class of natural choices, in the above sense.
These are characterized by having two-point functions whose leading
short-distance behavior is the same as in Minkowski space (Wald
1994).\footnote{This is a little bit of an oversimplification, as there
may also be infrared effects.}  Probably similar results are true for
other field equations.  In this paper, though, it will be unnecessary
to examine {\em how} $J$ is determined; it will be a datum.

Since the representation will have the same mathematical structure as
Fock space, we may speak of creation and annihilation operators.  In
general, these will have no simple interpretation in terms of particles,
but refer to some other fundamental modes (whichever physical modes
constitute the $+\i$ eigenspace of $J$).  We may also speak of a
``vacuum'' in this sense.  In general this ``vacuum'' state has
only mathematical interest, and does not have a physical interpretation
as the vacuum.  It will {\em not} be invariant under a change from one
representation to a unitarily equivalent one.

Now if $g:H\to H$ is a continuous linear map preserving $\omega$, then
$g$ induces a change $J\mapsto gJg^{-1}$, and so

\begin{corr}
A symmetry $g$ of $(H,\omega )$ is unitarily implementable on the
representation determined by $J$ iff $J-gJg^{-1}$ is Hilbert--Schmidt.
\end{corr}

\noindent This is simply the restatement, in the present formalism, of
the well-known criterion for Bogoliubov transformations to be
unitarily implementable.

Suppose now one has a one-parameter group $g(t)$ of motions of $H$
preserving $\omega$.  In most physical applications, this group is
{\em strongly continuous}, meaning that for any fixed $v\in H$, the
function $t\mapsto g(t)v$ is continuous.  Under these circumstances,
there is a generator $A$ so that $g(t)=e^{tA}$.  In applications, this
generator is typically a partial differential operator.  For example,
for evolution in time for the Klein--Gordon field, one has $H=\{
(\dot\phi ,\phi )\}$ and
\begin{equation}
 A=\left[
  \begin{array}{cc}
   0&1\\
  \Delta -m^2&0
  \end{array}\right]\, ,
\end{equation}
where $\Delta$ is the spatial Laplacian.

\subsection{Definitions and Notation}

Throughout, we shall let $H_\C$ be a complex infinite-dimensional
separable Hilbert space.  The complex inner product on $H_\C$ will be
denoted $\langle\cdot ,\cdot\rangle$.  We shall let $H$ be the underlying
real Hilbert space.  Then we write $J:H\to H$ for the real-linear map
given by $v\mapsto \i v$, and
\begin{eqnarray}
 (v,w)&=&\Re\langle v,w\rangle\label{ipef}\\
     \omega (v,w)&=&\Im \langle v,w\rangle
\end{eqnarray}
Then $(\cdot ,\cdot )$ is the canonical real inner product on $H$ and
$\omega$ is a symplectic form on $H$ which is non-degenerate in that it
defines isomorphisms from $H$ to its dual.  Note that
\begin{equation} (v,w)=\omega (v,Jw )\, .\end{equation}
Thus any two of $\omega$, $J$ and $(\cdot ,\cdot )$ determine the third.

Throughout, the {\em real } adjoint of a real-linear operator (perhaps
only densely defined) $L$ will be denoted $L^*$.  Thus the defining
relation is $(v,L^*w)=(Lv,w)$ with domain $D(L^*)=\{ w\in H\mid
(v,L^*w)=(Lv,w)$ for some $L^*w$ for all $v\in D(L)\}$.  
A useful result is the

\begin{proposition}
If $g:H\to H$ is a continuous linear map preserving $\omega$, then $g$ is
invertible and $g^{-1}=-Jg^*J$.  Conversely, the adjoint of $g$ is
$g^*=-Jg^{-1}J$.
\end{proposition}

\begin{proof} One has
\[ \omega (v,-Jg^*Jgw) =-(v,g^*Jgw)=-(gv,Jgw)=\omega (gv,gw) =\omega
(v,w)\]
for all $v,w\in H$, and similarly
\[ (v,g^*w)=(gv,w)=\omega (gv,Jw)=
  \omega (v,g^{-1}Jw)=-(v,Jg^{-1}Jw)\,
.\]
\end{proof}

\begin{definition}
The {\em symplectic group} of $H$ is
\[ \Sp (H)=\{ g:H\to H\mid g\mbox{ is linear, continuous
and preserves }\omega\}\,
.\]
Its elements are the {\em symplectomorphisms.}
\end{definition}

The symplectic group does not depend on the real inner product on $H$
(or on the complex structure); it depends only on $\omega$ and the
structure of $H$ as a Hilbertable space.  It has naturally the structure
of a Banach group, using the operator norm to define the topology.

\begin{definition}
The {\em restricted symplectic
group } of $H_\C$ is 
\[\rSp (H_\C )=\{g\in \Sp (H)\mid
  g^{-1}Jg-g\mbox{ is Hilbert--Schmidt}\}\, .\]
\end{definition}

That this set is closed under composition and inversion is a
consequence of the fact that the Hilbert--Schmidt operators form an
ideal.  There is a natural topology on $\rSp (H_\C )$; see Shale (1962).
(We shall not need this topology here, since we shall be concerned
exclusively with strong continuity.)
Note that the complex-linear and -antilinear parts of $g$ are
$g_\pm =(1/2)\bigl( g\mp JgJ\bigr)$.  Thus $g^{-1}Jg-g$ is
Hilbert--Schmidt iff $g_-$ is.

We recall that a {\em strongly continuous} one-parameter subgroup of
$\Sp (H)$ is a one-parameter subgroup $t\mapsto g(t)$ such that, for
each $v\in H$, the map $t\mapsto g(t)v$ is continuous.  (In general, one
can also consider {\em semigroups}, defined for $t\geq 0$, but as every
symplectomorphism is invertible, in our case every semigroup extends to
a group, which is strongly continuous iff the semigroup is.)  
According to the Hille--Yoshida--Phillips Theorem, such groups
have the form $g(t)=e^{tA}$, where $A$ is a densely-defined operator on
$H$ (with certain spectral properties), and $\| g(t)\| _{\rm op}\leq
Me^{|c|t}$ for some $M,c\geq 0$.  The spectrum of $A$ is confined to the
strip $| \Re\lambda |\leq c$.

We now wish to consider the action of $g(t)$ by conjugation on certain
operators.  It is not hard to see that, for any operator $L$ of finite
rank, the function $t\mapsto g(t)Lg(-t)$ is continuous in 
Hilbert--Schmidt 
norm.
Now suppose $L$ is a Hilbert--Schmidt 
operator.  Then for any $\epsilon >0$, we
may write $L=L_{\rm f}+L_\epsilon$, where $L_{\rm f}$ has finite rank
and $\| L_\epsilon\| _{\rm HS}<\epsilon$.  We then have
\begin{eqnarray}
\| g(t)Lg(-t)-L\| _{\rm HS}
  &\leq &\| g(t)L_{\rm f}g(-t)-L_{\rm f}\| _{\rm HS}
  +\| g(t)L_\epsilon g(-t) -L_\epsilon\| _{\rm HS}\nonumber\\
  &\leq&\| g(t)L_{\rm f}g(-t)-L_{\rm f}\| _{\rm HS}\\
  &&\qquad
  +\bigl( \| g(t)\| _{\rm op}\| g(-t)\| _{\rm op} +1\bigr)\epsilon
\end{eqnarray}
We may make this less than any given $\eta >0$, as follows.  Assume
first $|t|<1$.  Then the Hille--Yoshida--Phillips Theorem bounds $\|
g(t)\| _{\rm op}\| g(-t)\| _{\rm op}$, and hence choosing $\epsilon$ we
may make the second term as small as desired.  Then the first term may
be made small by restricting $|t|<\delta$.  Thus conjugation by $g(t)$
is continuous at the identity on the Hilbert--Schmidt operators.  But since
conjugation is a group action, we have proved

\begin{proposition}\label{ermengarde}
Let $g(t)$ be a strongly continuous one-parameter subgroup of $\Sp (H)$.
Conjugation by $g(t)$ induces a strongly-continuous
one--parameter group $G(t)=e^{t\cA}$ on the space of Hilbert--Schmidt 
operators
on $H$.
\end{proposition}

(The same proof would apply for the compact operators, or those of trace
class.)

\section{Characterization of Generators 
of $\rSp (H_\C )$\label{Characterization}}

In this section, we shall prove the main theorem.  We shall do this
in two main stages.  Our first goal is to find out what restrictions
must be made on $A$ in order that $\exp tA$ be unitarily implementable
for each $t$.  We shall then show that no additional hypotheses need be
made to ensure that these unitary maps can be chosen to form a
strongly-continuous one-parameter group.  Thus we shall be able to
conclude that, when $\exp tA$ is implementable for each $t$, a
Hamiltonian operator exists.  This final argument is delicate, and I do
not know of any reason {\em a priori} to expect the conclusion.

We begin by introducing a function measuring the change in complex
structure:  let
\begin{equation}
L(t) =g(t)Jg(-t)-J\, .
\end{equation}
The group law for $g(t)$ implies a ``twisted'' law
\begin{equation}
  L(s+t)=g(s)L(t)g(-s)+L(s)\, .
\end{equation}
If we formally differentiate this at $s=0$, we find the differential
equation
\begin{equation}
   L'(t)=AL(t)-L(t)A +L'(0)\, .
\end{equation}
Here 
\begin{equation}
L'(0)=AJ-JA\, .
\end{equation}
The differential equation has the formal solution
\begin{equation}
  L(t) =\int _0^t G(u)L'(0)\, \d u\, ,\label{cassandra}
\end{equation}
where
\begin{equation}
  G(t)Q=g(t)Qg(-t)
\end{equation}
for any operator $Q$.

We shall now show how these equations can be interpreted rigorously.

\begin{proposition}
$JA$ and $AJ$ are self-adjoint operators (on $H$ with the inner
product $(\cdot ,\cdot )$ defined in equation~(\ref{ipef}))
with domains $D(A)$ and
$JD(A)=D(AJ)$, respectively.
\end{proposition}

\begin{proof}
We have
\[ D((JA)^*)=\{ v\in H\mid (JAw,v)=(w,(JA)^*v)\mbox{ for all }w\in
D(JA)\}\, .\]
Now, if $w\in D(A)=D(JA)$, we have
\begin{eqnarray*}
  (JAw,v)&=&\partial _{t,0}\omega (Jg(t)w,Jv)\\
  &=&\partial _{t,0}\omega (w,g(-t)v)\\
  &=&-\partial _{t,0} (w,Jg(-t)v)\\
  &=&\partial _{t,0}(w,Jg(t)v)\, ,
\end{eqnarray*}
where $\partial _{t,0}$ means $\partial /\partial t$ evaluated at $t=0$.
The proof for $AJ$ is similar.

\end{proof}

The condition $D(AJ)=D(JA)$ is the condition that $D(A)$ be invariant
under $J$, that is, that the domain of $A$ be a complex space with
complex structure $J$.  This holds for all physical systems that have so
far been studied, but there are at least mathematical examples for which
it fails.  

\begin{corollary}
The commutator $[A,J]=AJ-JA$ is naturally defined as a form
\begin{equation} (v,[A,J]w)=(AJv,w)-(v,JAw)\end{equation}
on $D(AJ)\times D(A)$ (or on $D(A)\times D(AJ)$).  We have
\begin{equation} 
  (v,[A,J]w)=\partial _{t,0}(v,g(t)Jg(-t)w) =\partial _{t,0}(v,L(t)w)
\end{equation}
in this case.
\end{corollary}

\begin{proof}
We have
\begin{eqnarray*}
  (v,g(t)Jg(-t)w)&=&\omega (v,Jg(t)Jg(-t)w)\\
  &=&-\omega (g(-t)Jv,Jg(-t)w)\\
  &=&-(g(-t)Jv,g(-t)w)
\end{eqnarray*}
For $w\in D(A)$, we may write $g(-t)w=w-tAw+o(t)$; similarly,
the last line displayed above becomes
\[ -(Jv-tAJv+o(t),w-tAw+o(t))=-(Jv,w)+t(AJv,w)-t(Jv,AW)+o(t)\, ,\]
which has the desired derivative.

\end{proof}

\noindent And similarly:

\begin{corollary}
The $J$-antilinear part of $A$,
\[ A_-=(1/2)\bigl( A+JAJ\bigr)\, ,\]
exists as a $J$-symmetric form with domain $D(A)$, and also
with domain $JD(A)$.  The $J$-linear part of $A$,
\[ A_+=(1/2)\bigl( A-JAJ\bigr)\, ,\]
exists as a $J$-skew form on both of these domains.
\end{corollary}

We can now make sense of the integral formula for $L(t)$.  

\begin{proposition}
The formal expression (\ref{cassandra}) for $L(t)$ is valid
in the sense of forms on $JD(A)\times D(A)$.  That is, for
any $v\in JD(A)$ and $w\in D(A)$, one has
\[ (v,L(t)w)=\int _0^t\bigl\{
  (v,G(u)AJw) -(v,G(u)JAw)\bigr\} \d u\, .
\]
\end{proposition}

\begin{proof} 
This follows by integration by parts; one has only to note that
the terms can be arranged so that this is sensible.

\end{proof}

\subsection{Regularity of the Complex Structure}

We show here that the function $L(t)$, measuring the change in the
complex structure, is continuous and of at most exponential growth.

\begin{theorem}
If $g(t)$ is a strongly continuous one-parameter family of
restricted symplectic transformations, then
the function $t\mapsto L(t)$ is continuous (in the
Hilbert--Schmidt norm).
\end{theorem}

\begin{proof}
We shall first show that the function is measurable.
Let $\epsilon$ be a positive real number and $L_0$ a fixed
Hilbert--Schmidt operator.  The inverse image of the
ball of radius $\epsilon$ at $L_0$ is 
\[
\{ t\mid \sum _j \| \bigl( L(t)-L_0\bigr) e_j\|_{H}^2 <\epsilon ^2\}\, ,
\]
where $e_j$ form an orthonormal basis for $H$.  But the sum
is a pointwise-convergent 
sum of non-negative continuous functions, hence lower
semicontinuous, and hence measurable.

Now
for positive integers $n$, consider the sets $\{ t\in (-1,1)\mid \|
L(t)\| _{\rm HS}>n\}$.  These form a decreasing sequence of sets of
finite measure, with intersection $\emptyset$.  Thus for large enough
$n$ the set $S=\{ t\in (-1,1)\mid \| L(t)\| _{\rm HS}<n\}$ has positive
measure.  We may similarly find a set $T\subset (-1,1)$, symmetric about
the origin, of positive measure, and with $\| L(t)\| _{\rm HS}$ bounded
on $T$.  However then the set $\{ t_1+t_2\mid t_1,t_2\in T\}$ will
contain an interval about the origin, and from the ``twisted group law''
and the Hille--Yoshida--Phillips Theorem we have
\[ \| L(t_1+t_2)\| _{\rm HS}\leq \| g(t_2)\| _{\rm op}
    \| L(t_1)\| _{\rm HS} \| g(-t_2)\| _{\rm op} +\| L(t_2)\| _{\rm HS}
\]
is bounded on this interval.  Thus there exists an interval containing
the origin on which $\| L(t)\| _{\rm HS}$ is bounded.

We integrate the twisted group law in the form
\begin{equation} g(t)Jg(-t)-J=g(s+t)Jg(-s-t)-J+g(t)
  \bigl( g(s)Jg(-s)-J\bigr) g(-t)
\end{equation}
(for $t$ close enough to zero) over a closed interval $[s_0,s_1]$
near the origin:
\begin{eqnarray*}
(s_1-s_0)\bigl( g(t)Jg(-t)-J\bigr)
  &=& \int _{s_0}^{s_1}\bigl[ g(s+t)Jg(-s-t)-J\bigr] \, \d s\\
   && -g(t)\int _{s_0}^{s_1}\bigl[ g(s)Jg(-s)-J\bigr]\, \d s\, g(-t)\\
  &=& \int _{s_0+t}^{s_1+t}\bigl[ g(u)Jg(-u)-J\bigr] \, \d u\\
  &&  -g(t)\int _{s_0}^{s_1}\bigl[ g(s)Jg(-s)-J\bigr]\, \d s\, g(-t)
\end{eqnarray*}
The first term of the last line tends to $\int _{s_0}^{s_1}\bigl[
g(s)Jg(-s)-J\bigr]\, \d s$ as $t\to 0$.  The second term does, too, since
we have shown that conjugation by $g(t)$ is strongly continuous on the
Hilbert--Schmidt operators.  Thus $L(t)$ is continuous at the origin.

Finally, at any value of $t$, for small enough $s$, we have
\[ \| L(t+s)-L(t) \| _{\rm HS} =\| g(t)L(s)g(-t)\| _{\rm HS}
  \leq \| g(t)\| _{\rm op}\| L(s)\| _{\rm HS}\| g(-t)\| _{\rm op}\,
\]
which tends to zero as $s$ does.

\end{proof}

\begin{proposition}\label{dhritarashtra}
For strongly continuous one-parameter families of restricted
symplectic motions, one has
\[ \| L(t)\| _{\rm HS}\leq \alpha e^{\beta |t|}\]
for some $\alpha ,\beta\geq 0$.  (If $\| g(t)\| _{\rm op}\leq
Me^{c|t|}$, then one may choose any $\beta >2c$.)
\end{proposition}

\begin{proof}
A little algebra shows
\[ L(nt)=\bigl[ G((n-1)t+G((n-2)t)+\cdots +1\bigr] L(t)\, .\]
From the Hille--Yoshida--Phillips Theorem, we know $\| g(t)\| _{\rm
op}\leq Me^{c|t|}$ for some $M\geq 1$, $c\geq 0$.  Thus
\[ \| L(nt)\| _{\rm HS}\leq nM^2e^{2c|nt|} \| L(t)\| _{\rm HS}\, .\]
For any $u$ with $|u|\geq 1$, 
write $u=n+r$ where $n$ is an integer and $|r|<1$, where $r$
has the same sign as $n$.  Then we
have
\begin{eqnarray*}
  \| L(u)\| _{\rm HS}&=&\| L(n(1+r/n))\| _{\rm HS}\\
  &\leq& nM^2e^{2c|u|} \| L(1+r/n)\| _{\rm HS}\\
  &\leq& |u|M^2e^{2c|u|}\sup _{t\in (-2,1]\cup [1,2)}\| L(t)\| _{\rm HS}
\end{eqnarray*}
for $|u|\geq 1$.  The result now follows from elementary considerations.

\end{proof}

We are now in a position to establish a useful property of unitarily
implementable evolutions.

\begin{proposition}
Let $A$ be the generator of a strongly continuous one-par\-a\-meter
subgroup of the restricted symplectic group.  Then the $J$-antilinear
part $\left( (\lambda -A)^{-1}\right) _-$ of its resolvent $(\lambda -A)^{-1}$ 
is Hilbert--Schmidt for the real part of 
$\lambda$ sufficiently positive, and is $o(\lambda )$ in the 
Hilbert--Schmidt
topology as $\lambda\to +\infty$.
\end{proposition}

\begin{proof}
Using a subscript minus to denote $J$-antilinear parts, we have
\[ \bigl( (\lambda -A)^{-1}\bigr) _-
  =\int _0^\infty g_-(t)e^{-\lambda t}\, \d t\, .\]
{\em A priori}, this integral is known to exist only in the strong
sense.  However, it follows from the previous two results that
$g_-(t)=-(1/2)L(t)g(t)J$ is a locally integrable Hilbert Schmidt-valued
function and that the integral converges for the real part of $\lambda$
sufficiently positive.  Multiplying by $\lambda$, one easily shows that
the resulting integral tends to $g_-(0)=0$ as $\lambda\to +\infty$.

\end{proof}

The converse of this fails; one can easily create counterexamples by
considering direct sums of countably many two-dimensional symplectic
spaces.

We close this subsection with some further properties of $L$ and of
operators derived from it, which will be useful in what follows.

\begin{proposition}
$JL$ is a $J$-symmetric operator with spectrum strictly below unity.
\end{proposition}

\begin{proof}
We have
\[ JL = JgJg^{-1}+1= -(g^{-1})^*g^{-1} +1\]
where the asterisk denotes the real adjoint.

\end{proof}

Now let us put 
\begin{equation}
 g_\pm =(1/2)\bigl( g\mp JgJ\bigr)
\end{equation}
for the $J$-linear and $J$-antilinear parts of $g$.  We have
$JL=2g_-g^{-1}$, and hence $g_-g^{-1}$ is a $J$-symmetric operator
varying continuously in Hilbert--Schmidt norm, with spectrum bounded
strictly below $1/2$.

\begin{proposition}\label{fabian}
For strongly continuous one-parameter families of restricted symplectic
motions, the quantity
$gg_+^{-1} -1$ is a $J$-symmetric
Hilbert--Schmidt operator, varying continuously 
with $t$ in
Hilbert--Schmidt norm.
\end{proposition}

\begin{proof}
We have
\[ gg_+^{-1}-1=(g_-g^{-1})\bigl[ 1-g_-g^{-1}\bigr] ^{-1}\, .\]
Since the first factor varies continuously in Hilbert--Schmidt norm and
the second continuously in operator norm, the product varies
continuously in Hilbert--Schmidt norm.

\end{proof}

Note that $JL(-t)=Jg^{-1}Jg+1=-2Jg^{-1}g_-J$, so $g^{-1}g_-$ is also a
continuous $J$-symmetric Hilbert--Schmidt operator, as is
$g_+^{-1}g-1$.

\subsection{The Characterization Theorem}

\begin{theorem}\label{hermione}
A
generator $A$ of a strongly-continuous one-parameter subgroup of
$\Sp (H)$
is a generator of a strongly-continuous one-parameter subgroup of
$\rSp (H_{\C})$ iff the following condition holds:
Let $g(t)=e^{tA}$, and let $G(t)=e^{t\cA}$ be the associated
one-parameter group acting on the compact operators on $H$ by
conjugation.  Then for any $\lambda$ in the resolvent set of $\cA$, 
the quantity
$R(\lambda ,\cA )L'(0)$ 
(is defined as a limit in the space of linear forms on $D(AJ)\times D(A)$ and)
is a Hilbert--Schmidt operator.
\end{theorem}

\begin{proof}
Note that $D(AJ)\times D(A)$ is a Hilbertable space, and so will be the
space of linear forms on it.

We have $L(t)=\int _0^tG(u)L'(0)\, \d u$.  The idea will be to
integrate this against $e^{-\lambda t}$.  However, since $L'(0)$ is only
weakly defined, it will be easier to approach this integral as a
limit.  Consider then, for a Hilbert--Schmidt operator $B$,
\begin{eqnarray*}
 \int _0^\infty e^{-\lambda t}\int _0^t G(u) B\, \d u\, \d t
  &=&\int _0^\infty \lambda ^{-1}e^{(\cA -\lambda )u}B\, \d u\\
  &=&-\lambda ^{-1}R(\lambda ,\cA )\, B
\end{eqnarray*}
Here the spectrum of $\cA$ must lie in a strip $|\Re z|<2c$ for some
$c>0$, and so the integral converges for $\lambda$ sufficiently positive.
Now, as $B$ approaches $L'(0)$ (in the space of forms on 
$D(AJ)\times D(A)$), the left-hand side of this equation approaches
$\int _0^\infty e^{-\lambda t}L(t)\, \d t$ in the space of forms, but
this integral is in fact a Hilbert--Schmidt operator.  (This follows
from proposition~\ref{dhritarashtra}.)
Therefore, as $B$ tends to $L'(0)$ as a form, the
quantity $\lambda ^{-1}R(\lambda ,\cA )\, B$ tends to a limit, which we
denote $\lambda ^{-1}R(\lambda ,\cA )\, L'(0)$, equal to the integral.

For the converse, we note that
\[\int _0^t G(u)L'(0)\, \d u 
  =\lambda \int _0^tG(u)R(\lambda ,\cA )L'(0)\, \d u
  +(1-G(t))R(\lambda ,\cA )L'(0)\, ,\]
and the right-hand side is Hilbert--Schmidt in view of 
proposition~\ref{ermengarde}.

\end{proof}

This theorem is one of our main results.  In applications, the operator
$A$ is the Hamiltonian operator for the classical field equations, acting
on Cauchy data.  The complex structure $J$ is a pseudodifferential
operator whose singular part is determined by the local structure of the
field operator.  Thus $R(\lambda ,{\cal A})L'(0)$ can be computed by
pseudodifferential operator techniques from local data.  Whether it
exists as a Hilbert--Schmidt operator is in most cases easily read off
simply by considering the orders of the dominant terms.

\section{Existence of a Hamiltonian\label{Existence}}

The Characterization Theorem determines the conditions under which a
strongly-contin\-uous one-param\-e\-ter subgroup of $\Sp (H)$ lies
in $\rSp (H_{\C})$. 
In this case, it is what one might call {\em pointwise unitarily
implementable,} that is, for each $t$ there exists a unitary
transformation $U(t)$ on the Fock space implementing $e^{tA}$.  
(Each of these transformations is determined uniquely up to phase.)  One would
like to know if (with an appropriate choice of phases) we have
$U(t)=e^{\i t{\wH}}$ for a self-adjoint Hamiltonian ${\wH}$.
It turns out that this is always the case.

In order to prove this, and for its own interest, we shall work out an
explicit formula for $U(t)$.  I would expect that
formulas equivalent to this one (modulo
phase) are known.  However, it is worth going through the analysis
explicitly here, for two reasons.  First, the phase is of some interest
and is technically difficult to analyze.  Second, we need fine control
over some of the terms in order to establish the strong continuity of
$U(t)$, and so a careful presentation is worthwhile.

\subsection{The Representation\label{Representation}}

The representation (of the Weyl algebra of the field operators)
is defined as follows.  Let $Z^a\in H_\C$.  (We shall
use an index notation when convenient.)  The $Z^a$ will be creation
operators, with $\partial _a=\partial /\partial Z^a$ annihilation
operators.  Then the state wave functions are holomorphic functions
$\Psi (Z^a)$.  It is sometimes convenient to distinguish between these
wave functions and the abstract state vectors $|\Psi\rangle$; the two
are related by
\begin{equation}
 |\Psi\rangle =\Psi (Z^a)|0_Z\rangle\, ,
\end{equation}
where $|0_Z\rangle$ is the ``vacuum'' state.  In this equation, the
terms in the power series $\Psi (Z^a)$ are thought of as creation
operators.

The inner product is
\begin{equation}
\langle\Psi |\Phi\rangle =\int {\overline\Psi}\Phi
  e^{-\langle Z,Z\rangle}{\cal D}Z{\cal D}{\overline Z}\, .
\end{equation}
This integral is defined as the integral of $\overline\Psi \Phi$ against
a promeasure; alternatively, it may be regarded as a short-hand for
the power series in the coefficients of $\overline\Psi$, $\Phi$ it
formally determines.  The normalization is fixed so that the norm of
$\Phi (Z)=1$ is unity.

Now let $g=g(t)$ be a symplectomorphism.  It induces an action on the
field operators which is conventionally written as
\begin{eqnarray}
 {Z'}^a&=&\alpha ^a{}_bZ^b+\beta ^{ab}\partial _b\\
  {\partial '}_a&=&{\overline\alpha}_a{}^b\partial _b
    +{\overline\beta}_{ab}Z^b\, .
\end{eqnarray}
Here $\alpha$, $\beta$ are essentially the $J$-linear and -antilinear
parts of $g(t)$, and are known as Bogoliubov coefficients.  That $g(t)$
be a symplectomorphism is equivalent to
\begin{eqnarray}
  0&=&\beta ^{ac}\alpha ^b{}_c-\beta ^{bc}\alpha ^a{}_c\\
  \delta ^a{}_b&=& {\overline\alpha}_b{}^c\alpha ^a{}_c
  -{\overline\beta}_{bc}\beta ^{ac}\, .
\end{eqnarray}
Note that this implies $\alpha$ is invertible.  That $g(t)$ lie in
$\rSp$ is equivalent to requiring $\beta$ to be Hilbert--Schmidt.

The image of the vacuum is determined by the requirement that it be
annihilated by all operators ${\partial}'_a$, and from this one finds
the state is
\begin{equation}
 N\exp -(1/2)Q_{ab}Z^aZ^b
\end{equation}
(or more precisely $N\exp -(1/2)Q_{ab}Z^aZ^b |0_Z\rangle$), where
\begin{equation}
  Q_{ab}=\bigl({\overline\alpha}^{-1}\bigr) _a{}^c
  {\overline\beta}_{cb} 
\end{equation}
is symmetric and the normalization $N$ has modulus
\begin{equation}
  |N|=
  \bigl[\det\bigl( \delta ^a{}_b-{\overline Q}^{ac}Q_{cb}\bigr)
  \bigr] ^{1/2}\, .
\end{equation}
The conditions on $\alpha$ and $\beta$ above imply that this state is
well-defined (and that $|N|$, as defined here, is positive).

The evolution of a general state vector may now be determined.  Let is
write the abstract ket as
\begin{equation}
 |\Psi\rangle =\Psi (z)|0_Z\rangle =\Psi '(Z')|0_{Z'}\rangle\, ,
\end{equation}
where 
\begin{equation}
 |0_Z\rangle\, ,\quad
 |0_{Z'}\rangle =N\exp -(1/2)Q_{ab}Z^aZ^b|0_Z\rangle
\end{equation}
are the vacua with respect to $J$
and $g(t)Jg(-t)$.  Now write, in matrix notation,
\begin{eqnarray}
  {Z'}&=&\alpha Z+\beta\partial \\
  &=&\alpha Z+(\alpha\beta ^T)\partial _{\alpha Z}\\
  &=&e^{(1/2)\alpha\beta ^T\partial _{\alpha Z}\partial _{\alpha Z}}
  \alpha Ze^{-(1/2)\alpha\beta ^T\partial _{\alpha Z}\partial _{\alpha Z}}
 \, ,
\end{eqnarray}
where $\partial _{\alpha Z}=\partial /\partial (\alpha Z)$.
(Here we have used the relations $\partial /\partial (\alpha Z)
=\alpha ^{-1\, T}\partial _Z$ and $\alpha\beta ^T =\beta\alpha ^T$.)
The last line of the displayed equation is valid, for example,
as an operator identity with the exponentials defined by their formal
power series, acting on polynomials, and extends by linearity to
suitable holomorphic functions.
Then we have
\begin{eqnarray}
 \Psi (Z) &=&\Psi '(Z')Ne^{ -(1/2)Q_{ab}Z^aZ^b}\\
  &=&e^{(1/2)\alpha\beta ^T\partial _{\alpha Z}\partial _{\alpha Z}}
  \Psi '(\alpha Z)
  e^{-(1/2)\alpha\beta ^T\partial _{\alpha Z}\partial _{\alpha Z}}
  Ne^{ -(1/2)Q_{ab}Z^aZ^b}\, .
\end{eqnarray}

This formula defines $U(t)$, modulo phase.  We know that it is a
one-parameter projective unitary group.  If we can show that this
projective group is strongly continuous, and that the phases can be
chosen to make the full group strongly continuous, then we shall be
assured of the existence of a self-adjoint generator.

\subsection{Continuity of Some Operations}

\begin{proposition}
$Q_{ab}$ is a continuous function of $t$ in the Hilbert--Schmidt norm.
\end{proposition}

\begin{proof}
We shall work with ${\overline Q}^{ab}$, to avoid conjugating $\alpha$
and $\beta$.

For this, we must derive the precise relation between the $\alpha$'s,
$\beta$'s, and $g$.  This arises from the canonical quantization
prescription, which in our case amounts to the replacement of the
variables $Z^a$, ${\overline Z}_a$ with the operators $Z^a$, $\partial
_a$.  We see that $\alpha$ is precisely $g_+$, the $J$-linear part of
$g$.  We can work out $\beta$ from the identity
\[ {\overline V}_a{Z'}^a={\overline V}_a\alpha ^a{}_bZ^b
  +\beta ^{ab}{\overline V}_a{\overline Z}_b\, \]
from which we find
\[ \beta ^{ab}{\overline V}_a{\overline Z}_b
  = \langle V,gZ\rangle -\langle Z,g_+Z\rangle 
  =\langle V,g_-Z\rangle\]
and thus
\[ {\alpha ^{-1}}^a{}_b\beta ^{bc}{\overline Z}_a{\overline Z}_c
  =\langle Z,(g_+)^{-1}g_-Z\rangle\, .\]
It was shown in proposition~\ref{fabian} (and the comments following
that proposition) that $(g_+)^{-1}g_-$ is continuous in
Hilbert--Schmidt norm.

\end{proof}

\begin{proposition}
With the choice of phase $N=|N|$, the image of the $J$-vacuum varies
continuously with $t$.
\end{proposition}

\begin{proof}
Let $|\Psi _t\rangle$ be the image of the vacuum at time $t$.  Then
\begin{eqnarray*}
 \langle \Psi _t-\Psi _s| \Psi _t-\Psi _s\rangle
  &=& \int |N_t\exp -(1/2) Q_{tab}Z^aZ^b -N_s\exp -(1/2)Q_{sab}Z^aZ^b|^2
     \\ &&\qquad\times
     e^{-\langle Z,Z\rangle} {\cal D}Z{\cal D}\overline Z\\
  &=& 2-2N_sN_t\Re \det (I-{\overline Q}_sQ_t\bigr) ^{-1}\, .
\end{eqnarray*}
Since $Q$ varies continuously in Hilbert--Schmidt norm, for any fixed
$t$, this can be made as close to zero as desired by choosing $s$ close
enough to $t$.

\end{proof}

We now turn to a similar, more general computation.  The trigonometric
polynomials are dense in ${\cal H}$.  (This is the present formulation
of the well-known statement that the vacuum is a cyclic state for this
representation.)  We shall show that they vary continuously with $t$.

\begin{proposition}
With the choice of phase $N=|N|$, the image of any trigonometric
polynomial varies continuously with $t$.
\end{proposition}

\begin{proof}
It is enough to establish this for trigonometric monomials.

For any $A\in H$, let $W(A)=\exp\i ({\overline A}\cdot Z+A\cdot \partial
)$ be the corresponding Weyl operator.  
Here $A^\alpha$ is $A$ as an element of $H_\C$, and ${\overline A}_\alpha
=\overline{A^\alpha}$ is its complex conjugate.
Then a trigonometric monomial
is (a constant times) $W(A)|0\rangle$ for some $A$.  The image of this
state at time $t$ is
\begin{eqnarray*}
 &W\bigl( g(t)A\bigr) N_t\exp -(1/2)Q_{tab}Z^aZ^b =
 N_t\exp\left\{ -(1/2)\langle A_t,A_t\rangle \right.\\
  &\qquad\left. +(1/2)Q_{tab}A^a_t A^b_t
    -(1/2)Q_{tab}Z^aZ^b +\i ({\overline A}_{ta}-Q_{tab}A^b_t)Z^a
  \right\}\, ,
\end{eqnarray*}
where $A_t=g(t)A$.  We find
\begin{eqnarray*}
 \lefteqn{\langle\Psi _s |\Psi _t\rangle ={\overline N}_s N_t
  \det \left(\delta ^a{}_b-Q_{tbc}{\overline Q}_s^{ca}\right) ^{-1}}\\
&&\times
  \exp (1/2)\left\{ -\langle A_s,A_s\rangle -\langle A_t,A_t\rangle
  +Q_{tab}A_t^aA_t^b+\overline{Q _{sab}A_s^aA_s^b}\right\}\\
  &&\times
  \exp (1/2)
    \left[ \begin{array}{cc}
            {\overline A}_{ta}-Q_{tab}A_t^b
                      &A_s^a-{\overline Q}_s^{ab}{\overline A}_{sb}
           \end{array}\right]
    \left[ \begin{array}{cc}
            \delta ^a{}_b &{\overline Q}_s^{ab}\\
            Q_{tab}&\delta ^a{}_b
           \end{array}\right] ^{-1}
    \left[\begin{array}{c}
            A_s^a-{\overline Q}_s^{ab}{\overline A}_{sb}\\
            {\overline A}_{ta}-Q_{tab}A_t^b
           \end{array}\right] \, .
\end{eqnarray*}
Since, as $s$ approaches $t$, we have $A_s\to A_t$ and $Q_s$ to $Q_t$ in
Hilbert--Schmidt norm, this tends to unity as $s\to t$.

\end{proof}

\subsection{Existence Theorem for the Hamiltonian}

We are now in a position to prove the existence of the Hamiltonian.

\begin{theorem}\label{guillermo}
If $\e ^{tA}$ is a strongly-continuous one-par\-a\-meter subgroup of $\rSp$,
then there exists a self-adjoint operator $\wH$ on Hilbert space,
unique up to an additive constant, such that $U(t)=\e ^{\i t{\wH}}$
implements $\e ^{tA}$.
\end{theorem}

\begin{proof}
With the choice of phases $N=|N|$, we have a projective unitary 
representation
$U(t)$.  The cocycle representing its deviation from a true
representation is $U(s)U(t)U(-s-t)$.  This can be computed by lengthy
but straightforward means.  In matrix notation, we find it is
\[ \left(\det _\C (I+Q_t{\overline Q}_t)(I-Q_{s+t}{\overline Q}_s
  (I-Q_{t+s}{\overline Q}_{t+s})^{-1})\right) ^{1/2}\, .
\]
Here the quantity whose determinant is to be taken is of the form $I+T$,
where $T$ varies continuously in trace norm in $s$ and $t$.
From this it follows that the cocycle is continuous, and so a continuous
choice of phase is possible, making $U(t)$ into a one-parameter unitary
group.  Let such a choice be made.

Finally, we must show that this group is strongly continuous.  Since the
phases vary continuously, it is enough to show that the original,
projective representation is strongly continuous.  While this could
probably be done directly from the formula above, it is probably clearer
to give an indirect argument.  

In the previous subsection it was shown that 
$U(t)$ is strongly continuous on a dense family of
states.  
(Recall that now the phase has been chosen so that $U(t)$ is a
one--parameter unitary group.)
For any such state $|\Psi\rangle$ and any $t$, the state
$(1/t)\int _0^tU(u)|\Psi \rangle\, \d u$ is in the domain of $\partial
_{t,0}U(t)$.  It follows that $U(t)$ has a densely-defined generator,
which, because $U(t)$ is unitary, must be self-adjoint.

\end{proof}

There is an interesting consequence of the formulas above:

\begin{corollary}
We have
\[ \langle 0|e^{\i t{\hat H}}|0\rangle \not=0 \]
for every $t$.
\end{corollary}

\noindent This means that, for any linear unitarily implementable
evolution, if the state is initially vacuum, at any later time a quantum
measurement to determine the state will have a positive probability of
finding vacuum.  While this result would be trivial for the second quantization
of a one-particle Hamiltonian, here the operator $\hat{H}$ need not preserve
particle number, but may create or destroy pairs of particles, as the terms
with quadratic contributions in $Z$ or $\partial$ enter into the evolution.

We have shown that under certain conditions a self-adjoint Hamiltonian
$\H$ exists generating a one-parameter family $U(t)$ of unitary
transformations.  In the course of this argument, we determined $U(t)$
up to phase.  It would be of interest to determine the phase.
However, here I shall
only indicate that the problem is essentially one of renormalization.

We recall from the proof of theorem~\ref{guillermo} 
that the cocycle $U(s)U(t)U(-s-t)$ for
the projective representation was given by 
\begin{equation} 
 \left(\det _\C (I+Q_t{\overline Q}_t)(I-Q_{s+t}{\overline Q}_s
  (I-Q_{t+s}{\overline Q}_{t+s})^{-1})\right) ^{1/2}\, .
\end{equation}
After a little algebra, this can be rewritten as
\begin{equation}
 \left(\det _\C \alpha _{s+t}\alpha _s ^{-1}\alpha _t^{-1}\right) ^{1/2}\, .
\end{equation}
If the determinant of $\alpha$ were known to exist and depend
continuously on $t$,
it would be simple
to factor this:  the phase would simply be $(\det _\C\alpha )^{1/2}$.
However, the present hypotheses do not ensure that the determinant of
$\alpha$ is defined.  (We only know that $\alpha$ is the identity plus a
Hilbert--Schmidt term, not a trace-class term.)  Thus the isolation of
the phase is more delicate.  One can think of this as finding a
renormalized definition of $\det _\C\alpha$.

\section{Scalar Fields in Curved Space--Time\label{Examples}}

These papers were motivated by problems which arose in the theory of
quantum fields in curved space--time.  In this section, we apply
the theory to that case.  In particular, we settle an
outstanding question:  Is the Hamiltonian for such fields
self--adjointly implementable in generic circumstances?

There are already two sorts of evidence pointing to a negative answer
(Helfer 1996).  First, it is known that evolution in time by a finite
motion is not unitarily implementable.  One might think that in this
case a self-adjoint family of Hamiltonians could not exist, for if it
did one could integrate it to deduce a unitary evolution, which would
be a contradiction.  However, in the present, non-autonomous,
situation, the domains of the Hamiltonians could be time-dependent,
and so the integration might not be possible.  Thus the non-unitary
implementability of finite motions is not, in itself, enough to imply
non--self--adjoint implementability of the Hamiltonian.

The second sort of evidence comes from the formal expression for the
Hamiltonian.  This formal expression is known not to have any Hadamard
states in its domain.\footnote{I am using the term ``Hadamard state'' here to
mean a state vector $|\Psi\rangle$ in the Hadamard representation, rather than,
as is more common, a linear functional on the field algebra.  A more precise
definition will be given in the next footnote, after more terminology has been
developed.}
(Hadamard states are in a sense the nicest test
states in curved space--time, and often one takes as axioms that
certain operators should have well-defined actions on these states.)
While this is some indication of a pathological structure, it does not
prove that the Hamiltonian fails to exist --- the Hamiltonian could be
defined on some recondite domain, or its formal expression, derived
under the assumption of a certain renormalization prescription being
valid, might be incorrect.  Thus, the formal singularity of the
Hamiltonian is also inconclusive.

We shall show however that these arguments do suggest the correct
answer:  the Hamiltonian is not self-adjointly implementable.
These conclusions (and somewhat broader ones) could be deduced a bit
more quickly from the results of the next paper, but we wish to
illustrate how the general structure developed here applies.

The general set-up is the following.  We consider a space--time
$(M,g_{ab})$ which is oriented, time-oriented and globally
hyperbolic.  Global hyperbolicity ensures that relativistic field
equations are well-posed, and is necessary to ensure that a quantum
field theory can be constructed along conventional lines.  (See Wald
1994 for an outline of the construction of the quantum field theory.)
We shall also assume that the Cauchy surfaces are compact.  This is
only done for technical reasons (it rules out all infrared difficulties
and ambiguities):  the problems we shall uncover will be manifested in
the local, ultraviolet, divergences of certain traces, and would be
present in any Hadamard quantization, whether the Cauchy surfaces are
compact or not.

The field equation is
\begin{equation}
 (\nabla _a\nabla ^a +m^2)\phi =0\, ,
\end{equation}
and the symplectic form is $\omega (\phi ,\psi )=\int _\Sigma (\psi \,
{}^*d\phi -\phi\, {}^*d\psi )$.  Here $\Sigma$ is any Cauchy surface.

The complex structure is determined from
the Hadamard two-point function, and is a certain pseudodifferential
operator.  If we decompose the initial data for the field at $\Sigma$
as $\left[\begin{array}{c}\phi\\
\dot\phi\end{array}\right]$, as usual, and choose normal
coordinates (in terms of the induced metric) on $\Sigma$, then one has
for the symbol of $J$
\begin{equation}
  \left[\begin{array}{cc} {\rm sym}\, \alpha & |\xi |^{-1}\\
                         -|\xi |&-{\rm sym}\,\alpha\end{array}\right]
+ \left[\begin{array}{cc} O(|\xi |^{-2}) & O(|\xi |^{-3}\log |\xi |)\\
                         O(|\xi |^{-1}\log |\xi |&O(|\xi |^{-2})
      \end{array}\right]\, ,
\end{equation}
where 
\begin{equation}
 2{\rm sym}\, \alpha =\pi _a{}^a|\xi |^{-1}-\pi ^{ab}\xi _a\xi
_b|\xi |^{-3}\, ,\label{fahrnsnagel}
\end{equation}
with $\xi _a$ the Fourier transform variable,
its norm with respect to the three-metric on $\Sigma$ is $|\xi |$, and $\pi
_{ab}$ the second fundamental form of $\Sigma$ (Helfer 1996).  (This
$\alpha$ is not the same as the Bogoliubov coefficient.)\footnote{Once the
complex structure $J$ has been fixed, and the representation constructed as in
the previous section, the Hadamard states may be defined as follows.  They  are
the results of applying polynomials of creation operators $\omega (\psi
,{\widehat\phi}_{-})$ to the $J$-vacuum.  Here the creation operator
${\widehat\phi}_{-}$ is the $J$-negative frequency part of the field, and the
test functions $\psi$ are required to be smooth (with compactly supported
Cauchy data --- a requirement that is automatically fulfilled here).  The more
common notion of a Hadamard state as a linear functional on the field algebra
essentially corresponds to a density matrix formed from our Hadamard states.}

We shall consider, for simplicity, evolution along the unit timelike
normal to $\Sigma$.  In this case, the operator $A$ is
\begin{equation}
 \left[\begin{array}{cc} 0&1\\ -s^2&0\end{array}\right]\, ,
 \label{excellent}
\end{equation}
where we have put $s=\sqrt{-\Delta +m^2}$, with $\Delta$ the Laplacian
on the surface.  This operator is determined from the classical stress--energy
\begin{equation}
 T^{\rm classical}_{ab}=\nabla _a\phi\nabla _b\phi -(1/2)g_{ab}
   \nabla _c\phi\nabla ^c\phi
   \label{classstress}
\end{equation}
in the usual way.

It is convenient to make a change of basis to make $A$ diagonal. 
Accordingly, we shall put $\phi =\phi _{(+)} +\phi _{(-)}$ and $\dot\phi
=-\i s(\phi _{(+)}-\phi _{(-)})$.  Here $\phi _{(\pm )}$ are {\em not }
the positive- and negative-frequency parts of $\phi$ (which are
defined using $J$), but are the projections of $\phi$ onto the
eigenspaces of $A$.  Acting on $\left[\begin{array}{c}\phi _{(+)}\\ \phi
_{(-)}\end{array}\right]$, the operator $A$ is
\begin{equation}
 A=\left[\begin{array}{cc} -\i s&0\\ 0&\i s\end{array}\right]\, ,
\end{equation}
and the symbol of $J$ is
\begin{equation}
  {\rm sym}\, J=
  \left[\begin{array}{cc} -\i &\alpha\\ \alpha &\i\end{array}\right]
  +O(|\xi |^{-2}\log |\xi |)\, .
\end{equation}
From these equations, we may read off the symbol of $g(t)Jg(-t)$, as a
Fourier integral operator:
\begin{equation}\label{megalomania}
  \left[\begin{array}{cc} 
  -\i & {\rm sym}\,\alpha (x- \hat{\xi} t)\e ^{-2\i |\xi| t}\\
  {\rm sym}\,\alpha (x+\hat{\xi} t)\e ^{2\i |\xi |t}
  &\i\end{array}\right]\, ,
\end{equation}
where $\hat{\xi}$ is the unit vector in the direction $\xi$.  One sees
directly from this that $g(t)Jg(-t)-J$, for finite $t$, will generally be
an operator of order $-1$, and so not Hilbert--Schmidt.  This is a
direct estimate of $g(t)Jg(-t)-J$; it is not necessary here to use the
characterization theorem.

However, it is instructive to see the connection between this and the
characterization theorem.  
For this we need to compute $Q=(\lambda -{\cal
A})^{-1}{\cal A}J$.  Thus one should solve $[A,J]=\lambda
Q-[A,Q]$ for $Q$, which is a linear evolution equation for $Q$
along the vector field generating $A$.  This is an autonomous system,
because the coefficients $A$ and data $J$ are given (as operators on
initial data) at $\Sigma$.

Integration of this system can be accomplished directly, or by more
general formal means.  We shall take advantage of the computation for
$g(t)Jg(-t)$ we have already made.  We have the identity
\begin{equation}
  (\lambda -{\cal A})^{-1}{\cal A}J =\int _0^\infty \e ^{-\lambda t}
  \e ^{t{\cal A}}{\cal A}J\, \d t\, .
\end{equation}
Using the formula~(\ref{megalomania}), we find $(\lambda -{\cal A})^{-1}
{\cal A}J$ is
\begin{equation}
2\i |\xi |\left[\begin{array}{cc}
  0&-\int _0^\infty {\rm sym}\, \alpha (x -{\hat\xi}t)\e ^{-(\lambda +2\i
|\xi |)t}\, \d t\\
    \int _0^\infty {\rm sym}\, \alpha (x +{\hat\xi}t)\e ^{-(\lambda -2\i
|\xi |)t}\, \d t &0\end{array}\right]\, .
\end{equation}
For large enough real $\lambda$, the contributions from $\alpha$ occur
in arbitrarily small neighborhoods of $t=0$, and so the leading behavior
(of the upper-right term, say) is
\begin{equation}
  -2\i |\xi |({\rm sym}\, \alpha ) (\lambda +2\i |\xi |)^{-1}\, .
\end{equation}
The contribution of this term to the Hilbert--Schmidt norm is
\begin{equation}\label{hubbahubba}
 (2\pi )^{-3}\int _\Sigma \d \Sigma (x)\int _{|\xi |>\epsilon >0} \d ^3\xi\, 
  {{4|\xi |^2}\over{\lambda ^2 +4|\xi |^2}} | {\rm sym}\, \alpha |^2\, .
\end{equation}

We now do the angular part of the integral.  Let us put
${\rm sym}\,\alpha =\beta ^{ab}{\hat\xi}_a{\hat\xi}_b /|\xi |$
(cf. equation~(\ref{fahrnsnagel})).  Then
the angular contribution to the expression~(\ref{hubbahubba})
is (up to radially symmetric factors)
\begin{eqnarray}
 \int _{S^2}\beta ^{ab}\beta ^{cd}{\hat\xi}_a{\hat\xi}_b
  {\hat\xi}_c{\hat\xi}_d\d ^2{\hat\xi}
  &=&(4\pi /5)\beta ^{ab}\beta ^{cd} \eta _{(ab}\eta _{cd)}\\
  &=&(4\pi /15)\bigl( \beta ^a{}_a\beta ^b{}_b+2\beta ^{ab}\beta
_{ab}\bigr)\, ,
\end{eqnarray}
where $\eta _{ab}$ is the three-dimensional Euclidean metric and the
parentheses on the subscripts indicate symmetrization.
This is a positive-definite symmetric form in $\beta ^{ab}$, and so is
positive unless $\beta ^{ab}$ vanishes identically, that is, unless
${\rm sym}\,\alpha$ vanishes identically.

Turning to the radial integral, since
$\alpha$ is of order $-1$, this is ultraviolet divergent
unless $\alpha$ vanishes identically. 
Inspecting the form of ${\rm sym}\,\alpha$ 
(equation~\ref{fahrnsnagel}),
we see that this would
require $\pi _{ab}$ to vanish identically.  Thus we have proved:

\begin{theorem}\label{Schubert}
Let $(M,g_{ab})$ be an oriented, time-oriented, globally hyperbolic
space--time with compact Cauchy surfaces.  Consider the quantum field
theory of a scalar field subject to the equation
\[ (\nabla _a\nabla ^a+m^2)\phi =0\]
in a Hadamard representation.  Let $\Sigma$ be a particular Cauchy
surface, and $A$ the operator generating evolution along the unit
normal determined by the usual classical stress--energy,
equation~(\ref{classstress}).
If the second fundamental form of $\Sigma$ does not vanish
identically, then $A$ in not self-adjointly implementable.
\end{theorem}

In the set of Cauchy surfaces, those with vanishing second fundamental
forms constitute a thin set in any reasonable topology.  Indeed, the
class of globally hyperbolic space--times {\em admitting} a Cauchy
surface with vanishing second fundamental form is arguably a thin set
in any reasonable topology.  (In the case of zero classical
stress--energy, 
these are
space--times which possess time-reflection symmetry.)  We may say that
generically $A$ is not self-adjointly implementable.

More generally, one would conjecture that a
Hamiltonian $A$ corresponding to evolution along a vector field $v^a$
at $\Sigma$ would not be self-adjointly implementable unless $v^a$
satisfied Killing's equation (restricted to $\Sigma$) (cf. Helfer 1996,
p. L133).

We close with a comment about the conformally coupled massless field
\begin{equation}
  \left( \nabla _a\nabla ^a +(1/6)R\right)\phi =0
  \label{burgoo}
\end{equation}
with its ``new, improved'' stress--energy
\begin{equation}
\begin{split}
  T^{\rm classical}_{ab}&=(2/3)\nabla _a\phi\nabla _b\phi -(1/6)g_{ab}\nabla
  _c\phi\nabla ^c\phi -(1/3)\phi\nabla _a\nabla _b\phi \\
  &\quad +(1/18)R\phi ^2g_{ab}
    -(1/6)\phi ^2R_{ab}\, .\end{split}
    \label{improvedstress}
\end{equation}
The field equation here agrees with that of the massless case of the ordinary
scalar field in cases where $R=0$; in particular, in Minkowski space.  But the
evolution of the field data as generated by the stress--energy is different,
because the evolutions are determined by the Poisson brackets of the fields with
the energy integrals $\int T_{ab}\xi ^a\d \Sigma ^b$, and the different
stress--energies yield different integrals.
(One way of viewing this is that the ``new, improved'' stress--energy contains
terms which correct the evolution of $\dot\phi$ to account for $\phi$ having
conformal weight $-1$.)

One can analyze the conformally coupled field in a manner completely parallel to
that for the ordinary scalar field.  
I shall not give the details of the computations here.  One finds that
the operator $A$ 
is then no longer given by 
equation~(\ref{excellent}), but has correction terms which are
\begin{equation}
\left[\begin{array}{cc}
  0&0\\ 0&-(2/3)\pi _a{}^a
  \end{array}\right] +\mbox{lower-order terms}\, .
\end{equation}
These turn out to cancel the 
highest-order pure-trace contributions of $\pi _{ab}$ to
$g(t)Jg(-t)-J$, and we wind up with

\begin{theorem}
Let $(M,g_{ab})$ be an oriented, time--oriented, globally hyperbolic
space--time with compact Cauchy surfaces.  Consider the quantum field
theory of a conformally coupled scalar field (with field equation~\ref{burgoo})
in a Hadamard representation.  Let $\Sigma$ be a particular Cauchy
surface, and $A$ the operator generating evolution along the unit
normal determined by the ``new, improved'' stress--energy, 
equation~(\ref{improvedstress}).
If the second fundamental form of $\Sigma$ is not pure trace,
then $A$ in not self-adjointly implementable.
\end{theorem}

\section{Comments\label{Summary}}

The results of this paper were outlined in the introduction, and no
summary will be given here.  Rather, this section contains a few
technical comments.

The two main general results in this paper are Theorem~\ref{hermione}, which
establishes which classical Hamiltonian vector fields generate motions
which are implementable on the quantum Hilbert space by unitary
transformations, and Theorem~\ref{guillermo}, which shows that when such
unitary implementation is possible a quantum Hamiltonian necessarily
exists.  

The latter result is gratifying physically, in that it means a
certain type of pathology is absent.  (The pathology would be that each
classical canonical transformation in the one-parameter family would
have a unitary implementation, but that it would not be possible to
choose this family of unitary motions with strong enough continuity
properties to guarantee the existence of a self-adjoint generator.)
However, at least the present argument for this is rather delicate (one
has ``just enough'' convergence to establish it).  It would be
worthwhile to find a simple argument to replace it.

In some sense, the lesson of Theorem~\ref{hermione} is that what is
important is not so much the generator $A$ of the classical motions
(that is, the Hamiltonian vector field), as its Lie adjoint ${\cal
A}=[A,\cdot ]$:  one needs
\begin{equation}
  ({\cal A}-\lambda )^{-1}{\cal A} J
\end{equation}
to be Hilbert--Schmidt (for sufficiently large $\lambda$) in order that
$A$ be self-adjointly implementable.  
If ${\cal A}$ were known to have a spectral representation, then this criterion
would amount to saying that, in terms of its spectral resolution, the
quantity ${\cal A}J$ projected near the origin
(that is, $\int _{|\lambda |<a} \d { E}(\lambda ){\cal A}J$,
where $\d { E}(\lambda )$ is the spectral
measure) was Hilbert--Schmidt, and
that $J$ projected near infinity was.  In other words, the complex
structure $J$ should have certain asymptotics in terms of the spectral
resolution of ${\cal A}$.

We shall see in the sequel that classically positive Hamiltonians have
Hamiltonian vectors which are necessarily spectral operators
(and hence ${\cal A}$ is spectral).  However,
in more general circumstances, this need not be the case.
For example, take the phase space to be
the countable direct sum $\oplus _n \{
(p_n,q_n)\in\R ^2\}$ of two-dimensional phase spaces.  The Hamiltonian
function will be 
\begin{equation}
H=p_1q_2+p_2q_3+\cdots
\end{equation}
so that the induced canonical transformation is, in block form with
respect to the $(q,p)$ decomposition
\begin{equation}
\left[\begin{array}{ccccccccc}
 0&1&0&0&\cdots &&&&\\
 0&0&1&0&\cdots &&&&\\
 &&&\ddots &&&&\\ 
 &&&&0&0&0&0&\cdots\\
 &&&&-1&0&0&0&\cdots\\
 &&&&0&-1&0&0&\cdots \\
 &&&& &&&&\ddots\end{array}\right]
\end{equation}
(blank places are occupied by zeroes).  This decomposes into two shift
operators, which are the well-known to be non-spectral (Dunford \&
Schwartz 1971).  
(We remark that this operator is not only bounded but contractive:  $\|
Av\|\leq\| v\|$, where $\|\cdot\|$ is the standard $L_2$ norm.)

\paragraph{Acknowledgments}
It is a pleasure to thank members of the
University of Missouri--Columbia Mathematics Department, 
especially Peter Casazza, Nigel Kalton, Yuri
Latush\-kin and Stephen Mont\-gomery--Smith, for helpful
conversations; and Paul Robinson for a useful electronic exchange.

\end{document}